\documentclass[12pt]{article}
\usepackage{epsfig,amsfonts,amssymb}
\usepackage{hyperref}
\usepackage{cite}
\input epsf.sty
\usepackage{cite}

\newcommand{\hab}{}

\def\ZZZ{{\hbox{ Z\kern-1.6mm Z}}}
\def\RRR{{\hbox{ R\kern-2.4mm R}}}
\def\CCC{{\hbox{ C\kern-2.0mm C}}}
\def\zzz{{\hbox{z\kern-1mm z}}}

\newcommand{\ten}{{(10)}}
\newcommand{\bet}{{( b )}}

\newcommand{\qq}{k}
\newcommand{\pp}{l}
\newcommand{\nn}{\nonumber \\}

\newcommand{\vt}{\vartheta}

\newcommand{\vtau} {\vec \tau}
\newcommand{\vj} {\vec J}
\newcommand{\vxi} {\vec \xi}
\newcommand{\vu} {\vec u}
\newcommand{\htau} {\vec \eta}
\newcommand{\vc}{\vec\chi}
\newcommand{\vpsi} {\vec \psi}

\newcommand{\qeq}{{\hbox{=\kern-2.3mm ? \kern.5mm }}}
\renewcommand{\qeq}{=}

\newcommand{\rrho}{r}
\newcommand{\bA}{{\bf A}}
\newcommand{\tx}{\wt x}
\newcommand{\bG}{{\bf G}}
\newcommand{\bF}{{\bar F}}
\newcommand{\bbb}{{\bar b}}
\newcommand{\gam}{\tau}
\newcommand{\eps}{\epsilon}
\newcommand{\vareps}{\varepsilon}
\newcommand{\ra}{\rangle}
\newcommand{\la}{\langle}
\newcommand{\T}{\chi_{T}(k)}
\newcommand{\Tm}{\chi_{T}(k')}
\newcommand{\Cn}{{\cal C}_n}
\newcommand{\vp}{\varphi}
\newcommand{\ve}{\varepsilon}
\newcommand{\tl}{\lambda}
\newcommand{\dt}{(\vec \nabla T)^2}
\newcommand{\hp}{{\wh\Phi}}
\newcommand{\hq}{{\wh Q_B}}
\newcommand{\he}{{\wh\eta_0}}
\newcommand{\ha}{{\wh{A}}}
\newcommand{\lllb}{\Bigl\langle\Bigl\langle}
\newcommand{\rrrb}{\Bigr\rangle\Bigr\rangle}
\newcommand{\tf}{\wt f}
\newcommand{\sss}{{\cal L}_{av}}
\newcommand{\bx}{\bar x}
\newcommand{\bw}{\bar w}
\newcommand{\ws}{{\wt\sigma}}
\newcommand{\wrh}{{\wt\rho}}
\newcommand{\wv}{{\wt v}}

\newcommand{\vv} {\bar v}
\newcommand{\uu} {\bar u}
\newcommand{\K}{{\rm K_1}}
\newcommand{\Kt}{{\rm \widetilde K_1}}

\newcommand{\B}{b'}
\newcommand{\C}{c\,'}
\newcommand{\bB}{\bar b'}
\newcommand{\Bu}{B_{\vec u}}
\newcommand{\VV}{{\cal V}}
\newcommand{\BB}{{\cal B}}
\newcommand{\DD}{{\cal D}}
\newcommand{\BBB}{{\cal B}}
\newcommand{\II}{{\cal I}}
\newcommand{\AAA}{{\cal A}}
\newcommand{\GG}{{\cal G}}
\newcommand{\KK}{{\cal K}}
\newcommand{\fff}{{\bf f}}
\newcommand{\ccc}{{\bf c}}
\newcommand{\FF}{{\cal F}}
\newcommand{\JJ}{{\cal J}}
\newcommand{\HH}{{\cal H}}
\newcommand{\MM}{{\cal M}}
\newcommand{\CC}{{\cal C}}
\newcommand{\bC}{{\bf C}}
\newcommand{\OO}{{\cal O}}
\newcommand{\QQ}{{\cal Q}}
\newcommand{\PP}{{\cal P}}
\newcommand{\EE}{{\cal E}}
\newcommand{\LL}{{\cal L}}
 \newcommand{\rrr}{\rangle\rangle}
\newcommand{\half}{{1\over 2}}
\newcommand{\wt}{\widetilde}
\newcommand{\wh}{\widehat}
\newcommand{\wc}{\wt}
\newcommand{\wb}{\bar}
\newcommand{\RR}{{\cal R}}
\newcommand{\NN}{{\cal N}}
\newcommand{\TT}{{\cal T}}
\newcommand{\bg}{\bar g}
\newcommand{\ba}{\bar a}
\newcommand{\bc}{\bar c}
\newcommand{\bd}{\bar d}
\newcommand{\bb}{\bar b}
\newcommand{\bT}{\bar \Theta}
\newcommand{\SSS}{{\cal S}}
\newcommand{\tlx}{\left(\tilde \lambda ; X^0(0) \right)}
\newcommand{\al}{\alpha}

\newcommand{\tk}{\tilde \kappa}

\newcommand{\ppp}{\prime\prime}

\newcommand{\omk}{\omega_n(\vec k)}
\newcommand{\onk}{\omega^{(N)}_{\vec k_\perp}}
\newcommand{\tI}{\wt\II}
\newcommand{\hI}{\wh\II}
\newcommand{\nI}{\II}

\newcommand{\cp}{\check\Phi}
\newcommand{\cps}{\Psi}
\newcommand{\crh}{\check\rho}
\newcommand{\cs}{\check\sigma}
\newcommand{\cv}{\check v}
\newcommand{\com}{\check\Omega}

\newcommand{\be}{\begin{equation}}
\newcommand{\ee}{\end{equation}}
\newcommand{\ben}{\begin{eqnarray}\displaystyle}
\newcommand{\een}{\end{eqnarray}}

\newcommand{\refb}[1]{(\ref{#1})}
\newcommand{\p}{\partial}
\newcommand{\sectiono}[1]{\section{#1}\setcounter{equation}{0}}
\newcommand{\subsectiono}[1]{\subsection{#1}\setcounter{equation}{0}}

\newcommand{\zet}{\zeta}

\newcommand{\gsim}{\stackrel{>}{\sim}}
\newcommand{\lsim}{\stackrel{<}{\sim}}

\newcommand{\Lamb}{\Lambda}

\def\one{{\hbox{ 1\kern-.8mm l}}}
\def\zero{{\hbox{ 0\kern-1.5mm 0}}}

\def\wa{{\wh a}}
\def\wb{{\wh b}}
\def\wc{{\wh c}}
\def\wdd{{\wh d}}

\renewcommand{\theequation}{\arabic{equation}}

\newcommand{\bea}[1]{\begin{eqnarray}\label{#1} }
\newcommand{\eea}{\end{eqnarray}}

\newcommand{\wJ}{\wt J}
\newcommand{\bN}{{\bf N}}

\newcommand{\aaa}{b}

\def\figaa{

\def\JPicScale{0.4}
\ifx\JPicScale\undefined\def\JPicScale{1}\fi
\unitlength \JPicScale mm

}

\begin{document}


\baselineskip 24pt

\begin{center}
{\Large \bf  Ultraviolet and Infrared Divergences in Superstring Theory}

\end{center}

\vskip .6cm

\vspace*{4.0ex}

\baselineskip=18pt

\centerline{\large \rm Ashoke Sen}

\vspace*{1.0ex}

\centerline{\large \it Harish-Chandra Research Institute}
\centerline{\large \it  Chhatnag Road, Jhusi,
Allahabad 211019, India}

\vspace*{1.0ex}
\centerline{\small E-mail:  sen@mri.ernet.in}

\vspace*{1.0ex}

\centerline{\bf Abstract}  \bigskip

Superstring theory is known to be free from ultraviolet divergences but suffers from the
usual infrared divergences that occur in quantum field theories.
After briefly reviewing 
the origin of ultraviolet finiteness of superstring theory we describe recent progress 
towards
the understanding of infrared divergences in superstring theory. 

\bigskip

\begin{center}
(Invited article to the KIAS Newsletter, 2015)
\end{center}

\bigskip

\vskip .5in

\vfill \eject

\baselineskip=17.6pt




Quantum field theory is the standard theoretical 
tool for studying the physics of elementary particles. The most commonly used
approach for studying quantum field theories is perturbation theory, where we
take all the interaction effects to be small and carry out a Taylor series expansion
 in the coupling constants
-- the parameters that label  the interaction strengths --
of various physical quantities like the scattering amplitudes. 
The coefficients of the Taylor series expansion are given by sum of
{\it Feynman diagrams}, each of which represents an integral
over certain number of {\it loop momenta}. For a quantum
field theory in $d$ space-time dimensions a
typical integral takes the form
\be \label{exy1}
\int d^d \ell_1 \cdots d^d l_g \, \prod_{j=1}^r (k_j^2 + m_j^2)^{-1} \, \NN
\ee
where each $\ell_i$ is a $d$-dimensional vector labelling loop momenta,
each $k_j$ is a $d$-dimensional vector given by appropriate linear combination
of the $\ell_i$'s and the momenta $p_1, \cdots p_n$ carried by  the incoming and
outgoing
particles whose scattering amplitude we are trying to calculate, $m_j$
denotes the mass of one of the particles in the theory and the
{\it numerator factor} $\NN$ is a polynomial in the loop momenta $\{\ell_i\}$ and
the external momenta $\{p_k\}$. The components of the vector $k_j$ (and
similarly for $\ell_j$) are labelled as $(k_j^0,\cdots k_j^{d-1})$, and
$k_j^2\equiv -(k_j^0)^2 + (k_j^1)^2+\cdots + (k_j^{d-1})^2$.
The number of $\ell_i$'s and $k_j$'s, the expressions
for the $k_j$'s in terms of the $\ell_i$'s and the $p_k$'s, which mass $m_j$
to use in a given factor in \refb{exy1} and the precise expression for the
numerator factor $\NN$ are all fixed by the Feynman rules for a given Feynman 
diagram. Therefore all one needs to do to compute the scattering amplitude is to
carry out the integrals of the form given in \refb{exy1} and add the contributions
from different diagrams. An expression containing integration
over $g$ loop momenta, like the one appearing in \refb{exy1}, is usually referred to as
a $g$-loop contribution to the amplitude.

This has been an enormously
successful program and lies at the heart of most of what we know about
elementary particles. However, quantum field theories have inherent divergences -- infinities
encountered in the evaluation of \refb{exy1} -- which need 
to be dealt with before we can make concrete predictions. 
These divergences can be broadly classified into two kinds -- {\it ultraviolet} and
{\it infrared}.
The ultraviolet divergences come from the region of integration where one or more
of the $\ell_i$'s in \refb{exy1} become large. The  
infrared divergences arise from the vanishing of one or more factors of $(k_j^2+m_j^2)$.
It turns out that the ultraviolet divergences are unphysical, and can be removed in a class
of quantum field theories called renormalizable quantum field theories. For describing the
theory of elementary particles we use this kind of quantum field theories. On the other hand
the infrared divergences have physical origin, in that their appearance signals that we are
not asking the right question. Once the right question is asked, these divergences automatically
disappear. Typical examples of infrared divergences are {\it tadpole divergences}
which arise when we incorrectly identify the ground state of the system about
which we carry out the perturbation
expansion, and {\it mass renormalization divergences} which arise when we
use the wrong mass of the external states for computing the scattering
amplitudes. These divergences typically arise when the ground state
and/or the masses of the particles change after taking into account the 
effect of interactions, and we are not careful in taking into account the effect
of these changes in our calculation.

In order to characterize these divergences it is useful to use the so called 
{\it Schwinger parameter}
representation for the {\it propagator}
factors $(k_j^2 + m_j^2)^{-1}$. For each propagator we introduce a real
parameter $s_j$ and write
\be \label{exy2}
(k_j^2 + m_j^2)^{-1}=\int_0^\infty ds_j\, \exp[-s_j (k_j^2 + m_j^2)]\, .
\ee
With the help of this identity, \refb{exy1} can be written as
\be \label{exy3}
\int_0^\infty ds_1 \cdots \int_0^\infty ds_r 
\, \int d^d \ell_1 \cdots d^d l_g \, \exp\left[-\sum_j s_j(k_j^2+m_j^2)
\right] \, \NN\, .
\ee
Since each $k_j$ is a linear combination of the $\ell_i$'s, the exponent is quadratic function
of the $\ell_i$'s for fixed $s_j$. Since the numerator factor $\NN$ is polynomial in $\ell_j$,
we can now explicitly perform the integration over the $\ell_j$'s using the standard rules of
Gaussian integration over multiple variables.
Special care is needed to
treat the integration over the $\ell_j^0$'s; due to the fact that $(k_j^0)^2$ appears
with a negative coefficient in the expression for $k_j^2$,
the coefficients of $(\ell_i^0)^2$ in the argument of the
exponential in \refb{exy3} is positive and the $\ell_i^0$ integrals are
{\it a priori} divergent. This is 
circumvented by the standard procedure of analytically continuing these
integrals so that the $\ell_i^0$ integrals run along the imaginary axis.
Once this is done, one can carry out the integration over the $\ell_i$'s 
without encountering any divergence, and express \refb{exy3} as
\be \label{exy4}
\int_0^\infty ds_1 \cdots \int_0^\infty ds_r 
\,  F(\{s_i\})\, ,
\ee
for some function $F$ of the $s_j$'s. 
It is easy to verify that the ultraviolet divergences of 
the original integral, coming from
the region of large $\ell_j$, now will appear as a divergence in the integral \refb{exy4} from the
region where a subset of the $s_j$'s go to zero. On the other hand infrared divergences of
the original integral will appear in \refb{exy4} from the region where one or more $s_j$'s become
large.

In superstring theory (which for our discussion will stand for four different varieties
of string theory named as
{\it SO(32) heterotic}, {\it $E_8\times E_8$ heterotic}, {\it type
IIA} and {\it type IIB} string theories) 
we replace the notion that the elementary
building blocks of matter are point particles by the notion that they are strings -- one dimensional
extended objects. 
The main motivation for superstring theory stems from the fact that this theory
automatically incorporates gravity in its framework. This is to be contrasted with
quantum field theory, which has great difficulty in incorporating gravity in its fold.
The naive quantum field theory that one gets by
applying the usual rules of quantum field theory to general theory of relativity
--  the theory developed by Einstein a hundred years ago -- leads to a 
non-renormalizable theory and has uncontrolled ultraviolet divergences.

Superstring theory comes with its own prescription for 
computing scattering amplitudes which seems to differ from the sum over Feynman diagram 
expansion that emerges from a quantum field theory. 
The intrinsic difference arises from the fact that whereas particle trajectories are
described by curves in space-time, the trajectory of a string is described by a 
surface in space-time -- often referred to as the {\it world-sheet}. 
This intuitive picture allows us to represent the 
string scattering
amplitudes as integrals over the {\it moduli space} of two dimensional 
Riemann surfaces -- the moduli space being a space
whose different points label different Riemann surfaces.
In particular the $g$-loop, $n$-point scattering
amplitude in superstring theory is given by an expression
of the form:
\be \label{exy6}
\int_{\MM_{g,n}} \prod_{i=1}^{6g-6+2n} d m_i \, f(\{m_j\})\, .
\ee
Here $\MM_{g,n}$ denotes a $6g-6+2n$ dimensional 
moduli space of {\it genus} $g$ Riemann
surface with $n$ marked points -- the genus of a Riemann surface being the
number of handles that the Riemann surface has.
The integrand $f(\{m_j\})$ is given by the correlation function
of certain operators in a two dimensional conformal field theory inserted at the marked points
on the Riemann surface.
Which conformal field theory to use is determined by the specific background around 
which we study the superstring theory, whereas which operators in the conformal field theory we 
should use for our calculation is determined by the states whose scattering 
amplitude we want to compute.

This way of computing scattering amplitude {\it a priori} looks very 
different from the Feynman diagram expressions that we obtain from quantum 
field theories. However a closer examination reveals that in appropriate limit, the
integral over moduli space that comes from superstring theory begins to resemble the 
integrals over the Schwinger parameters $s_j$ in \refb{exy4} that appear in 
quantum field theories. Since the divergences in quantum field theory arise from
the integration over the parameters $s_j$, we see that to examine
the fate of those divergences in superstring theory  we have to examine possible divergences
arising from the integral over the moduli spaces of Riemann surfaces.

\begin{figure}
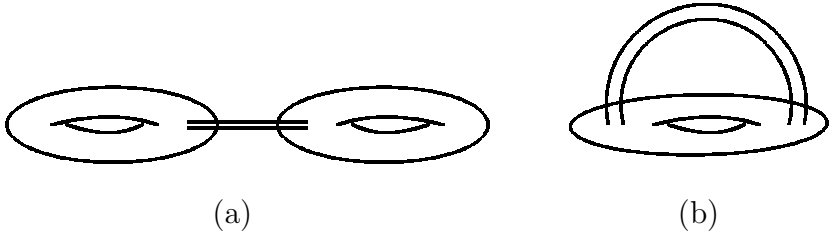


\begin{center}

\figaa

\end{center}

\vskip -.5in

\caption{Degeneration of Riemann surfaces. \label{fone}}

\end{figure}

Now the divergences from the integral over the moduli spaces of Riemann surfaces 
typically arise from the regions where the Riemann surface becomes singular. 
Study of singular Riemann surfaces is a well-developed subject, and all such
singularities are known to occur from {\it degenerations} of Riemann surfaces, where
the Riemann surface either becomes
a pair of Riemann surfaces connected by an infinitely
narrow tube (Fig.~\ref{fone}(a)), or develops an infinitely  narrow handle connecting two points
on a single Riemann surface (Fig.~\ref{fone}(b)).
By using the relation 
between the parameters $m_j$ labelling the moduli space of Riemann surfaces and the
parameters $s_j$ appearing in the expression for an amplitude \refb{exy4}
in quantum field theory, 
one can show that all such singularities in the moduli space of Riemann surfaces 
can be interpreted as the region where one or more of the $s_j$'s become infinite. 
Therefore
we conclude that all divergences in superstring theory can be 
interpreted as infrared divergences.

There is however a caveat. Even though the divergences coming from
singularities of the moduli space of Riemann
surfaces can be interpreted as infrared divergences, one may wonder 
whether there can be
divergences from the regular regions in the interior of the moduli space. 
This would happen
if the correlation function $f(\{m_j\})$ in the conformal field theory that we have to
compute blows up at some regular point in the interior of the moduli space. 
One does not
expect this to happen for a unitary conformal field theory, but it turns out that such a
singularity could arise from the correlation function in the non-unitary conformal
field theory of {\it superconformal ghost fields} -- the
fields which arise in the process of gauge fixing the 
supersymmetry transformation in the
superstring world-sheet theory\cite{Verlinde}. 
Physically these {\it spurious singularities} reflect the breakdown of
the gauge fixing procedure. Recently a completely systematic 
procedure for avoiding these
singularities has been developed\cite{1408.0571,1504.00609}. 
Furthermore one finds
that while the procedure itself is not unique, different ways of avoiding the
singularities lead to the same result for the scattering amplitudes.
This establishes that there are no
divergences coming from the interior of the moduli space, and the only possible divergences that
arise in superstring theory are infrared divergences. An alternative approach to
this problem, based on integration over the moduli space of {\it super Riemann 
surfaces},
has also been developed\cite{1209.5461}.

Therefore the relevant question is: how can we deal with the infrared divergences of superstring theory?
As we have already mentioned earlier, in a quantum field theory, infrared divergences have
physical origin; they reflect that we are asking the wrong questions. Experience with quantum field
theory also teaches us how to ask the right question and remove the infrared 
divergences. For example, quantum field theories have a systematic procedure for
taking into account possible changes in the ground state and/or masses of elementary
particles due to interaction, -- this is lacking in the conventional approach to
superstring perturbation theory.
Therefore,
if we had a quantum field theory whose Feynman diagrams reproduced the scattering 
amplitudes of superstring theory, then we would automatically 
know how to ask the right questions and avoid the infrared
divergences in superstring theory.

This is another area where there has been progress in recent years. It turns out that it is indeed
possible to write down a quantum field theory whose Feynman rules reproduce the amplitudes
of the form \refb{exy6} that come from superstring theory\cite{1508.05387}. 
This quantum field theory -- known as superstring field theory --  is somewhat
unusual, involving infinite number of fields and non-local interaction terms. Nevertheless it has
the structure inherent to a quantum field theory that allows us to remove the infrared divergences
exactly as we would do in a conventional quantum field theory.

To summarize, we now have a formulation of superstring theory which gives results
free from all divergences, infrared
and ultraviolet, when we ask the right questions.  The scattering amplitudes computed from this
formulation satisfy many of the desired properties {\it e.g.\ 
Ward identities} associated with
general coordinate transformations and other gauge symmetries\cite{1508.02481}. 
Work is in progress towards
proving other desired properties of the scattering amplitude -- {\it e.g.} unitarity (conservation
of probability) --  using this approach. We hope to make progress on this
front in the near future.


\small

\begin{thebibliography}{99}

\bibitem{Verlinde}
  E.~P.~Verlinde and H.~L.~Verlinde,
  ``Multiloop Calculations in Covariant Superstring Theory,''
  Phys.\ Lett.\ B {\bf 192} (1987) 95.
  doi:10.1016/0370-2693(87)91148-8

\bibitem{1408.0571}
  A.~Sen,
  ``Off-shell Amplitudes in Superstring Theory,''
  Fortsch.\ Phys.\  {\bf 63} (2015) 149
  doi:10.1002/prop.201500002
  [arXiv:1408.0571 [hep-th]].

\bibitem{1504.00609}
  A.~Sen and E.~Witten,
  ``Filling the gaps with PCOÕs,''
  JHEP {\bf 1509} (2015) 004
  doi:10.1007/JHEP09(2015)004
  [arXiv:1504.00609 [hep-th]].

\bibitem{1209.5461} 
  E.~Witten,
  ``Superstring Perturbation Theory Revisited,''
  arXiv:1209.5461 [hep-th].

\bibitem{1508.05387}
  A.~Sen,
  ``BV Master Action for Heterotic and Type II String Field Theories,''
  arXiv:1508.05387 [hep-th].

\bibitem{1508.02481}
  A.~Sen,
  ``Supersymmetry Restoration in Superstring Perturbation Theory,''
  arXiv:1508.02481 [hep-th].

\end{thebibliography}
\end{document}